\documentclass[desactivate]{aa}

\usepackage{graphicx}

\usepackage{natbib,twoopt}
\usepackage[pdftex,breaklinks=true]{hyperref} 
\bibpunct{(}{)}{;}{a}{}{,} 
\makeatletter
\newcommandtwoopt{\citeads}[3][][]{\href{http://adsabs.harvard.edu/abs/#3}%
{\def\hyper@linkstart##1##2{}%
\let\hyper@linkend\@empty\citealp[#1][#2]{#3}}}
\newcommandtwoopt{\citepads}[3][][]{\href{http://adsabs.harvard.edu/abs/#3}%
{\def\hyper@linkstart##1##2{}%
\let\hyper@linkend\@empty\citep[#1][#2]{#3}}}
\newcommandtwoopt{\citetads}[3][][]{\href{http://adsabs.harvard.edu/abs/#3}%
{\def\hyper@linkstart##1##2{}%
\let\hyper@linkend\@empty\citet[#1][#2]{#3}}}
\newcommandtwoopt{\citeyearads}[3][][]%
{\href{http://adsabs.harvard.edu/abs/#3}
{\def\hyper@linkstart##1##2{}%
\let\hyper@linkend\@empty\citeyear[#1][#2]{#3}}}
\makeatother

\usepackage[varg]{txfonts}

\newcommand{\modest}{\mbox{MODEST}}
\usepackage{url}
\usepackage{etoolbox}
\usepackage{xcolor}
\usepackage{orcidlink} 
\usepackage{xcolor}
\definecolor{darkblue}{cmyk}{55,17,0,0}
\hypersetup{colorlinks=true,
  linkcolor=darkblue,
  linkbordercolor=darkblue,
  citecolor=darkblue,
  urlcolor=black
  }

\begin{document}

\title{Magnetic properties of orphan penumbrae}

\author{ J.~S. {Castellanos~Dur\'an}\inst{1*} \orcidlink{0000-0003-4319-2009}
\and B. {L\"optien} \inst{1}
\and A. {Korpi-Lagg}\inst{1,2} \orcidlink{0000-0003-1459-7074} 
\and S.~K. {Solanki}\inst{1} \orcidlink{0000-0002-3418-8449}
\and M. {van~Noort}\inst{1} }

\institute{Max-Planck-Institut f\"ur Sonnensystemforschung, Justus-von-Liebig-Weg 3, 37077 G\"ottingen, Germany\\
$^*$email: castellanos@mps.mpg.de
\and Department of Computer Science, Aalto University, PO Box 15400, FI-00076 Aalto, Finland}

\date{Received 21 May 2025 / Accepted 10 July 2025}

\abstract {Orphan penumbrae (OPU) are features resembling sunspot penumbrae, but are not connected to an umbra. Here we compare OPUs and sunspot penumbrae, including their filaments. We also identify and describe the main mechanisms for the formation of OPUs and we characterise their decay process. Our study is based on spectropolarimetric inversions of active regions observed with the Hinode spectropolarimeter. We manually identified 80 individual OPUs, allowing us to study them statistically. In addition, we analysed the time-evolution of selected OPUs using data provided by the Helioseismic and Magnetic Imager. Orphan penumbrae display a broad range of shapes, associated with typically $\Omega$--shaped magnetic field configurations, where opposite polarity fields predominate at the two ends of the OPU. In addition, the properties of the OPU filaments are remarkably uniform between different OPUs, resembling the ones in sunspot penumbrae. Most OPUs form by either a patch of a penumbra separating from a sunspot, or by new magnetic flux emerging close to the polarity inversion line of an active region. We observe chromospheric fibrils above almost all OPUs in Hinode H$\alpha$ images, indicating that a part of the magnetic field of the OPUs extends to the chromosphere. Our results show that OPU filaments can form given a broad range of boundary conditions for the magnetic field. 
}


\maketitle

\section{Introduction}

Sunspots consist of a dark umbra and a comparatively brighter penumbra. The magnetic field in the umbra is rather uniform and is closer to being vertical than horizontal. In the penumbra, the magnetic field exhibits a complex structure. Regions with more vertical magnetic fields (spines) are interlaced with penumbral filaments (intra-spines), in which the magnetic field is close to horizontal \citep{1993ApJ...403..780T,1993A&A...275..283S,1993ApJ...418..928L,2013A&A...557A..25T}. The penumbral filaments are believed to be a manifestation of overturning magnetoconvection \citep{2008ApJ...677L.149S,2008A&A...488L..17Z,2009ApJ...691..640R,2011ApJ...729....5R, Scharmer2011Sci...EF,Joshi2011ApJ...EFdownflow,  2014ApJ...785...90R}. Penumbral filaments are highly elongated and  can be divided into three parts: a head, a body, and a tail \citep{2013A&A...557A..25T}. The head lies closest to the umbra, is bright in continuum intensity, exhibiting an upflow and a strong, almost vertical magnetic field of umbral polarity. It is followed by the body of the penumbral filament, in which the magnetic field is close to horizontal and which harbours a horizontal flow along its central axis. Furthest from the umbra lies the tail, which shows a strong downflow and an almost vertical magnetic field with the polarity being opposite to the one of the umbra. The flow observed running along penumbral filaments is known as the Evershed flow \citep{Evershed1909MNRAS..69..454E}.

In some cases, active regions (ARs) with sunspots harbour structures that resemble sunspot penumbrae, but which are not connected to an umbra. These regions are referred to as orphan penumbrae \citep[OPUs;][]{1991AdSpR..11e.225Z}. Similar to sunspot penumbrae, OPUs consist of penumbral filaments and exhibit an Evershed flow. However, the magnetic field is more uniform in an orphan penumbra, being close to horizontal and showing no clear indications of spines \citep{2014A&A...564A..91J}. Most OPUs are located close to the polarity inversion line (PIL) of the AR \citep{2012A&A...539A.131K,Kuckein2012A&A...FilamentOPUII,2014A&A...564A..91J,2014ApJ...787...57Z}.

The formation of sunspot penumbrae is not fully understood yet. Generally, it is attributed to the presence of an inclined magnetic field. However, it is not clear yet, which mechanism could lead to such inclined fields. Orphan penumbrae offer another perspective on both, the properties of magnetoconvection within penumbral filaments and the physical processes responsible for the formation of penumbrae. 

There is growing evidence that the formation of the penumbra is related to pre-existing magnetic fields in the chromosphere, such as the canopy of sunspots \citep{1980SoPh...68...49G,1982SoPh...79..267G,1992A&A...263..339S,Solanki1994A&A...Infrared-lines..Evershed,1999A&A...347L..27S}. These fields could act as a barrier, preventing newly emerging flux from rising up to higher atmospheric layers, thus keeping it strongly inclined \citep{1998ApJ...507..454L,2013ApJ...769L..18L}. Alternatively, the magnetic canopy of a pore might sink down and turn into the penumbra \citep{2012ApJ...747L..18S,2013ApJ...771L...3R,2014ApJ...784...10R,2016ApJ...825...75M,2020ApJ...899..129R}. 
See also \citet{Lindner2023A&A..PenumbraCanopy}. Numerical simulations also indicate that the formation of the penumbra is strongly connected to the magnetic field in the upper photospheric or chromospheric layers, since the extent of the penumbra strongly depends on the details of the top boundary conditions for the magnetic field \citep{2012ApJ...750...62R,2020A&A...638A..28J}. Simulations have also shown how convection can drag field from an overlying magnetic canopy down into the photosphere \citep{Pietarila2011ApJ...CanopyInternetwork}.

Similarly, there is no clear picture yet of the formation of OPUs. Several studies reported a connection between OPUs and overlying magnetic fields. \citet{2012A&A...539A.131K} observed several OPUs that were located underneath an AR filament. They interpreted these OPUs to be the photospheric counterpart of extremely low-lying chromospheric filaments. A connection between OPUs and chromospheric fibrils\footnote{The solar community uses the word filament in several contexts. This word is used for the photospheric penumbral filaments, as well as for the projection onto the disk of prominences, i.e., large-scale chromospheric filaments. In the paper, we use fibrils to denote thin, loop-like chromospheric filaments emanating from OPUs (and normal penumbrae) to avoid confusion.} was also noticed by \citet{2016A&A...589A..31B}. They found several OPUs to decay by ascending into the chromosphere once the overlying AR chromospheric filament had disappeared. In other cases, OPUs seem to form through the emergence of magnetic flux underneath a magnetic canopy \citep{2013ApJ...769L..18L,2014ApJ...786L..22G,2014ApJ...787...57Z}.  However, there are also examples where the formation of an OPU does not seem to be related to an overlying magnetic field at first glance. \citet{2014ApJ...787...57Z} also reported an OPU that formed by separating from the penumbra of a sunspot.

The magnetic field above OPUs appears to play a role not only during their formation of some OPUs but also during their decay. \citet{2014A&A...564A..91J}  
reported that the decay of an OPU relates to the submergence of a flat $\Omega$-shaped rope that trapped the flux emergence.

There seem to be several possible mechanisms for the formation of OPUs. Also, it is unclear how frequently these different mechanisms occur. The studies cited above only rely on individual OPUs or a small sample of OPUs. Altogether, the formation process has only been investigated for a handful of OPUs. In this paper, we perform a statistical analysis of a large sample of OPUs. Here we make use of the \modest{} database \citep{CastellanosDuran2024...modest}, which contains spectropolarimetric inversions of Hinode data for a large sample of ARs. We evaluate how much the general properties of OPUs and the properties of their filaments vary between different OPUs and how they differ from sunspot penumbrae. In addition, the large sample size allows us to identify and characterize the main mechanisms responsible for the formation of OPUs and to describe their decay process.

\section{Data and methods}
Our study makes use of the \modest{} catalogue, which consists of spectropolarimetric inversions of ARs. The \modest{} catalogue is an ongoing project. When we started our work on OPUs, the catalogue consisted of 710 inverted spectropolarimetric scans of 110~ARs. The catalogue is based on observations with the spectropolarimeter on the Solar Optical Telescope \citep[SOT-SP;][]{2007SoPh..243....3K,2008SoPh..249..167T,2008SoPh..249..233I,2013SoPh..283..579L} onboard the Hinode spacecraft. Hinode/SOT-SP performs spectropolarimetric scans using the \ion{Fe}{I} line pair at $6301.5$\,\AA \ and $6302.5$\,\AA. Most of the scans (682 scans) used for the catalogue were acquired in the so-called fast mode, with a spatial sampling of $0.32\arcsec$ per pixel. 

The \modest{} catalogue consists of the height-dependent atmospheric parameters of the ARs. These were derived from the Hinode/SOT-SP scans using the spatially coupled version of the SPINOR code \citep{2000A&A...358.1109F,2012A&A...548A...5V,2013A&A...557A..24V}, which uses the STOPRO routines \citep{Solanki1987PhDT} to carry out radiative transfer computations in local thermodynamic equilibrium. Here, three nodes in optical depth were used, placed at $\log \tau = -2.0,-0.8,0$. We note that we did not account for the $180^\circ$ azimuthal ambiguity and we did not transform the magnetic field to the local reference frame. In the following, the inclination of the magnetic field $\gamma$ is expressed with respect to the line-of-sight (LOS). We study the time-evolution of selected OPUs using data provided by the Helioseismic and Magnetic Imager \citep[HMI;][]{2012SoPh..275..229S} onboard the Solar Dynamics Observatory \citep[SDO;][]{2012SoPh..275....3P}. In addition, we observe the chromosphere above OPUs using the H$\alpha$ images obtained by Hinode/SOT-FG.

The total number of OPUs found and studied in this investigation is 80. These were identified and selected according to the criteria given in the next section.

\begin{figure*}[htbp]
\centering
\includegraphics[width=.94\textwidth]{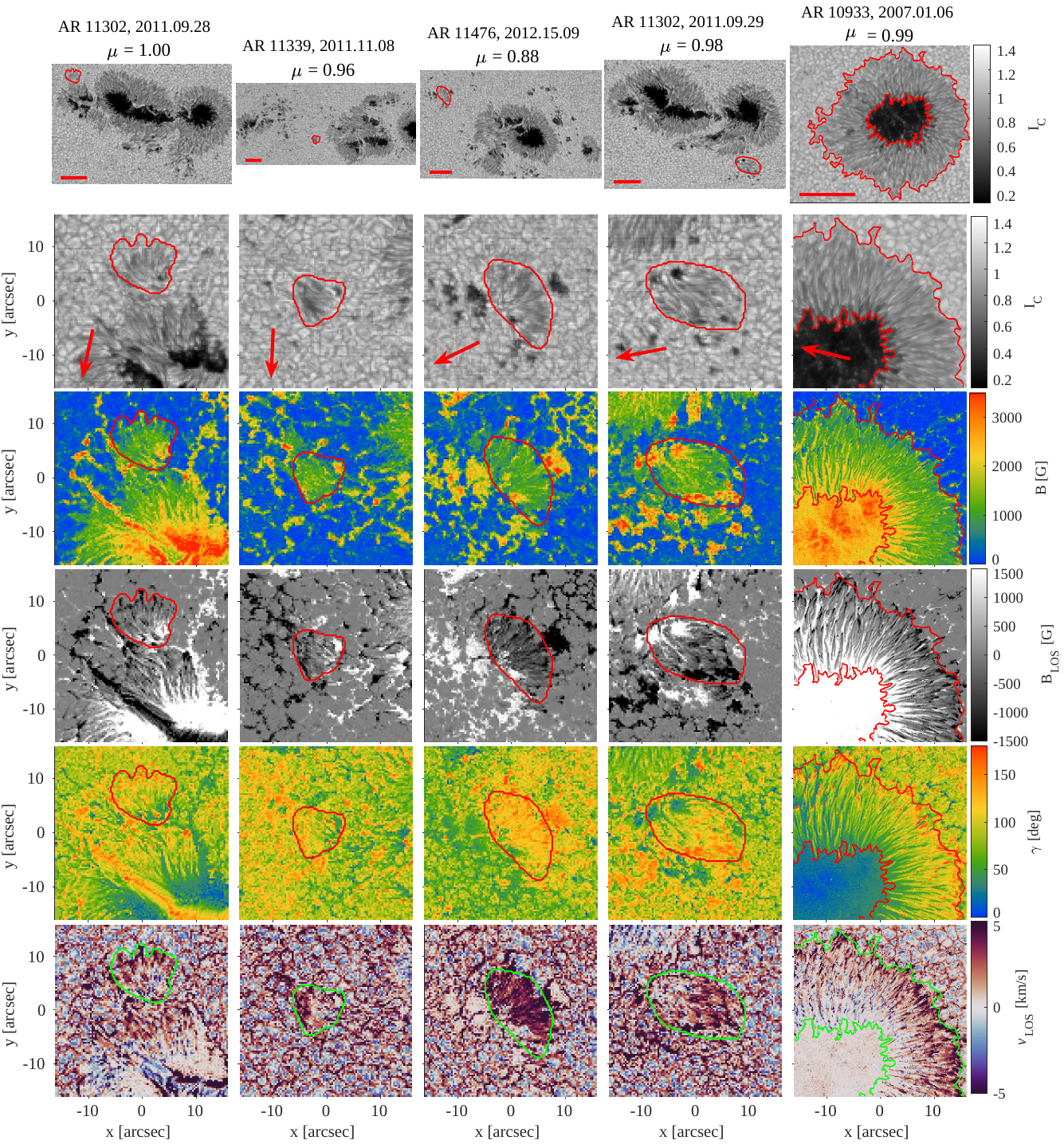}
\caption{Observables of selected OPUs. The four columns on the left show different examples of OPUs. The rightmost column shows the penumbra of a sunspot for comparison (AR\,10933 observed on 6 January 2007). The {\it top row} shows continuum intensity maps ($I_{\rm c}$) covering using the full FOV of the Hinode/SOT-SP scan. The rows below cover a smaller FOV of $32\arcsec \times 32\arcsec$ centred on the individual OPUs (or sunspot penumbra in the rightmost column). Rows 2 to 6 show the $I_{\rm c}$ of the region of interest, strength of the magnetic field $B$, longitudinal component of the magnetic field $B_{\rm LOS}$, inclination of the magnetic field $\gamma$ relative to the line-of-sight, and the LOS velocity $v_{\rm LOS}$, respectively. In all panels, the contours indicate the OPUs (or sunspot penumbra in the rightmost column). We adjusted the dominant polarity of $B_{\rm LOS}$ to be positive in the individual OPUs for better comparison. We show the magnetic field information and the velocity at $\tau = 1$. The red horizontal lines in the top row indicate a length of $20\arcsec$ and the arrows in the second-row point towards disk centre.}
\label{fig:OPU_examples}
\end{figure*}

\section{Comparison between orphan penumbrae and sunspot penumbrae}

Features resembling sunspot penumbrae can appear in a broad range of configurations, from simple, isolated OPUs to more complicated features, which are connected to other structures, such as sunspots or pores. Here we focus on OPUs which are isolated and are not connected to any other structures visible in continuum images. In total, we could identify 80 individual OPUs in the \modest{} catalogue, which fulfil these criteria. Some examples are shown in Fig.~\ref{fig:OPU_examples}, columns 1 to 4. As can be seen, even these simple OPUs exhibit a variety of shapes. The rightmost column of Fig.~\ref{fig:OPU_examples} shows a patch of the penumbra of a sunspot (AR\,10933 observed on 6~January 2007) for comparison.

Many OPUs are located very close to a sunspot (e.g., the OPU in AR\,11302 in columns 1 and 4 of Fig.~\ref{fig:OPU_examples}). When they are further away from a sunspot, the OPUs are usually situated close to the polarity inversion line (PIL) of the AR (e.g., the OPUs in AR\,11339 and in AR\,11476 shown in columns 2 and 3 of Fig.~\ref{fig:OPU_examples}). Most OPUs are dominated by one polarity of the magnetic field, even when they are close to the PIL. This is even the case for the OPU in AR\,11476, which is placed between two pores of opposite polarities. Only a few OPUs include a significant amount of flux of both polarities (e.g., the OPU in AR\,11302 in column 4 of Fig.~\ref{fig:OPU_examples}). The temporal evolution seen by SDO/HMI shows that this OPU was a region of new magnetic flux emergence. Initially, the emerging flux formed small pores that did not form a spot, rapidly turning into penumbrae but without umbrae. 

To confirm that the polarity of the OPUs are not seemingly reversed due to projection effects, we analysed SDO/HMI magnetograms when the ARs hosting the OPUs crossed the central meridional, where projection effects are lowest. This check was done for those ARs where the location of the OPU allowed it. This test confirmed that the statement about the magnetic polarity of OPUs being dominated by one polarity still holds.

Orphan penumbrae share a lot of common properties and resemble sunspot penumbrae in many aspects. For almost all OPUs, the OPU filaments are aligned in one direction, either lying seemingly parallel to each other (e.g. column 4 of Fig.~\ref{fig:OPU_examples}), or ordered in a fan shape (e.g. columns 1 and 3 of Fig.~\ref{fig:OPU_examples}). The heads (which exhibit an upflow) of OPU filaments tend to be located on one side of the OPU, and the tails (which harbour downflows and magnetic fields of opposite polarity) are situated on the opposite side of the OPU. OPUs in which some of the filaments appear parallel aligned but the head and tails are swapped, displaying oppositely directed flows exist, but they are rare  \citep[][see also Fig~\ref{fig:OPU_Halpha}]{2014A&A...564A..91J}. These results also imply that the Evershed flow in OPUs also occurs from the heads to the tails of filaments, as in normal penumbrae. For those OPU filaments displaying a reversed flow, the tails (heads) lie on the opposite side of the OPU compared to the others.

However, there are also differences between OPUs and sunspot penumbrae. Unlike sunspots, OPUs do not exhibit spines in between their filaments \citep[cf.][]{2014A&A...564A..91J}. In addition, no OPUs have been reported (and none are found in our sample) that harbour more than one row of filaments, while the penumbrae of large sunspots are much more extended than the length of individual penumbral filaments.

\begin{figure}[htbp]
\centering
\includegraphics[ width=0.5\textwidth]{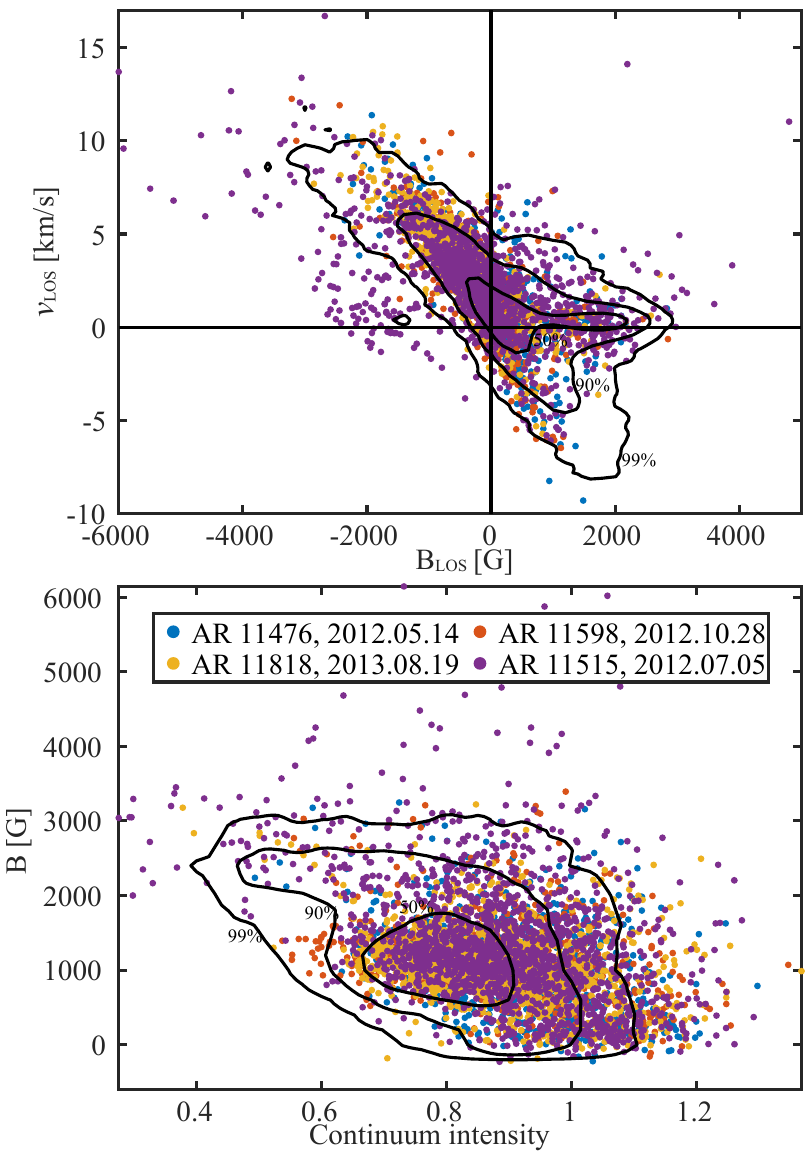}
\caption{Scatter plots between different observables for various OPUs and comparison with the penumbra of a sunspot. {\it Top:} scatter plot between $v_{\rm LOS}$ and $B_{\rm LOS}$. {\it Bottom:} $B$ vs. $I_{\rm c}$. The different colours correspond to the individual OPUs shown in Fig.~\ref{fig:OPU_examples} (indicated by the red contours in that figure). The black contours represent the 2D kernel density estimates of the respective observables for the penumbra of the sunspot shown in the rightmost column of Fig.~\ref{fig:OPU_examples}, with $99\%$, $90\%$, and $50\%$ of the distribution being within the respective contours. We adjusted the $B_{\rm LOS}$ polarity associated with the heads of filaments to be positive in the individual OPUs for a better comparison. The scatter plot in the top panel only includes pixels in which $B > 500$\,G to exclude surrounding patches of plage.}
\label{fig:OPU_example_scatter}
\end{figure}

In the next step, we evaluate whether the properties of the OPU filaments themselves vary between the different OPUs and whether they differ from those within sunspots. The most straightforward way to do this would be to derive an ensemble average of the OPU filaments in each OPU, as was done by \citet{2013A&A...557A..25T} for a sunspot. Unfortunately, individual filaments cannot be identified within the OPUs in a straightforward manner. This is because there are no prominent spines in OPUs, and so, the contrast between neighbouring OPU filaments is too low to be resolved in the Hinode/SP fast mode data. We can only infer the properties of the OPU filaments in an indirect way using scatter plots between different observables within individual OPUs (see Fig.~\ref{fig:OPU_example_scatter}). The scatter plot between $v_{\rm LOS}$ and $B_{\rm LOS}$ (top row in Fig.~\ref{fig:OPU_example_scatter}) shows that the OPUs and the penumbra of the sunspot contain up- and downflows of opposite polarity, as is expected for the heads and tails of penumbral filaments. For a better comparison, the $B_{\rm LOS}$ polarity associated with the filament heads was adjusted to be positive in the individual OPUs. Also, to minimise possible projection effects, only those OPUs were chosen that were observed when the ARs were crossing the central meridional. Most OPUs in this figure have similar distributions of $v_{\rm LOS}$ and $B_{\rm LOS}$ and resemble the penumbra of the sunspot. The main difference between the sunspot penumbra and the OPUs is that the penumbra covers a larger range of both $B_z$ and LOS velocity than the OPUs (the points belonging to the OPUs) do not fill the black contours that outline the locations of the sunspot penumbral data points. 

Only the bipolar OPU in AR\,11302 has a small patch, in which the downflows and the magnetic field are particularly strong (First and fourth columns of Fig.\ref{fig:OPU_examples}). Projection effects can be ruled out as the OPU in AR\,11302 was close to disk centre.  The pixels in the OPUs, which do not exhibit strong LOS velocities but strong magnetic fields correspond to small pore-like structures in the OPUs, close to the heads of the penumbral filaments, and to spines in the case of the penumbra of the sunspot. AR\,11302 contains pore-like features of both polarities. At the time this region was observed, it has fully emerged.

The bottom panel in Fig.~\ref{fig:OPU_example_scatter} shows a scatter plot between $B$ and $I_{\rm c}$. This figure allows separating between the pores and spines (which are dark and which exhibit strong magnetic fields) and the body of the filaments (which are brighter and harbour weaker $B$). Again, OPU and sunspot penumbrae show comparable properties \citep[cf.][]{Mathew2007A&A...spotcontrast}. However, the continuum intensity $I_{\rm c}$ of the OPU filaments varies between the different OPUs. Also, for all OPUs, the average continuum intensity in the OPU filaments is higher than that in the penumbra of the sunspot, basically because points with lower intensity are missing in the OPUs. There are also a significant number of points in the OPUs that are brighter than 99\%\ of all pixels in the sunspot penumbra.  In sunspots, the brightness of the penumbral filaments depends on the size of the spot, with smaller sunspots harbouring brighter penumbral filaments \citep{2020A&A...639A.106L,2021A&A...655A..61L}. Orphan penumbrae are smaller than sunspots, so their enhanced brightness could be related to their size. It is unclear, though, why the $I_{\rm c}$ of the OPU filaments differs between the OPUs. Besides the fact that OPUs miss the darkest pixels found in sunspot penumbrae, the points belonging to OPUs tend to cluster at the lower and intermediate field strength and are underrepresented at the higher field strengths found in sunspot penumbrae.

\section{Formation of orphan penumbrae}

\subsection{Identification of the major formation processes of orphan penumbrae}
In this section, we identify the major processes which lead to the formation of OPUs. We infer the mechanisms for the formation of the OPUs detected first in the \modest{} catalogue using data from HMI. HMI offers continuous observations with a high cadence, allowing us to study the time evolution of the OPUs in detail. Unfortunately, this is not possible for all OPUs in the \modest{} sample, because SDO was only launched in 2010, four years after the launch of Hinode. In addition, many OPUs form on the far side of the Sun and are already fully evolved when they reach the FOV of HMI. Nonetheless, we could capture the formation of 57 OPUs using HMI data.

\subsection{Separation from a sunspot}
\begin{figure*}
\centering
\includegraphics[width=.85\textwidth]{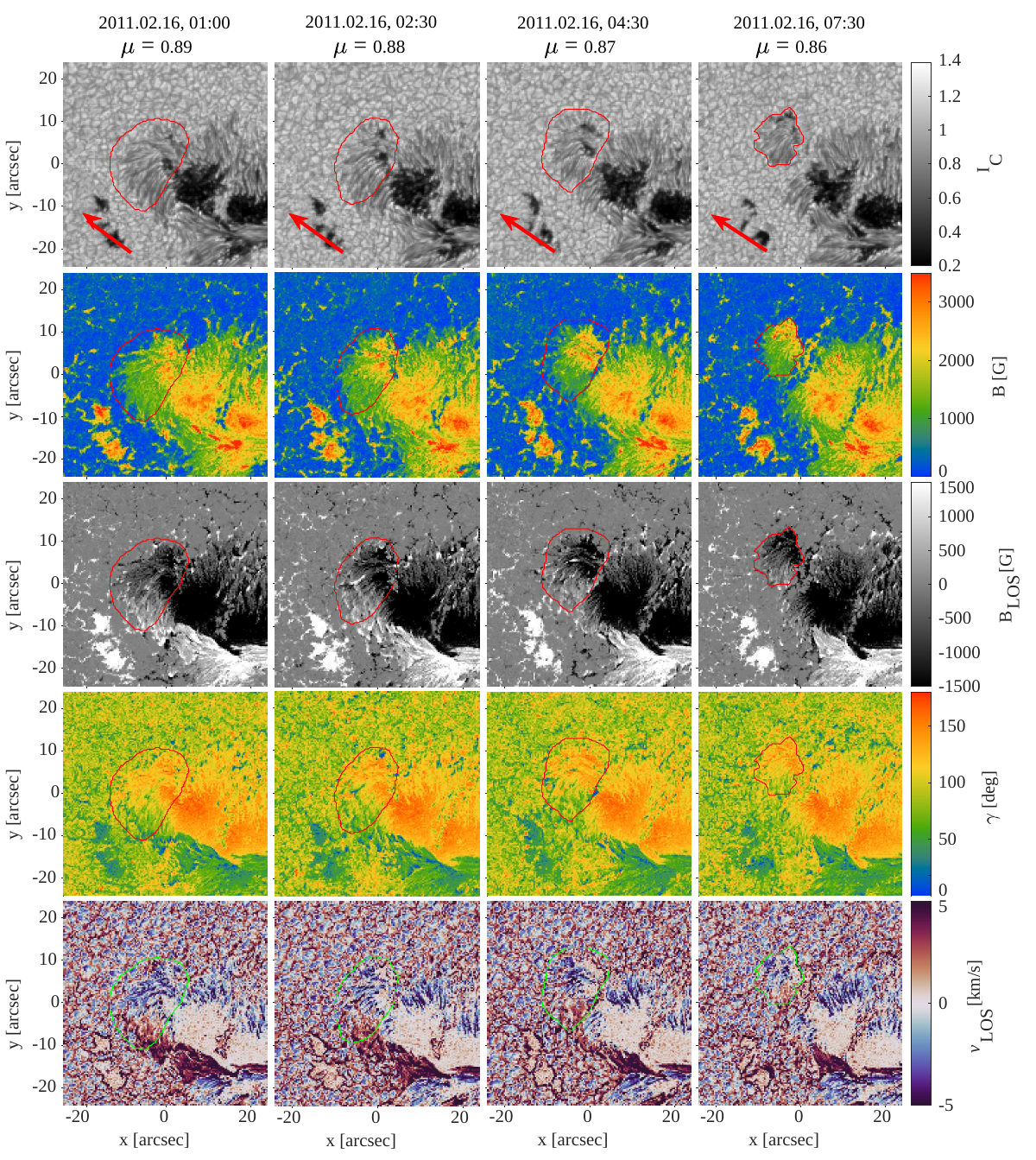}
\caption{Time-series showing the formation of an OPU in AR\,11158. The individual columns show the observables of the OPU at different times, from 1:00 UT to 7:30 UT on 16 February 2011. {\it From top to bottom:} $I_{\rm c}$, $B$, $B_{\rm LOS}$, $\gamma$, and $v_{\rm LOS}$, i.e. the same quantities in the same order as in Fig.~\ref{fig:OPU_examples}. All panels have a FOV of $48\arcsec \times 48\arcsec$ and the contours indicate the orphan penumbra. This OPU forms by separating from a sunspot. We show the magnetic field and the velocity at $\tau = 1$. The arrows in the top row point towards disk centre.}
\label{fig:OPU_separation}
\end{figure*}

We identified two major mechanisms for the formation of OPUs with a roughly equal occurrence rate. These mechanisms are the separation from the penumbra of a sunspot (26 OPUs) and the emergence of new flux (24 OPUs). The formation process of the remaining seven OPUs in the sample is less clear, because of the presence of multiple sunspots or pores. For some OPUs, the formation process is captured by a time series of several Hinode/SOT-SP scans, allowing us to study the two main processes in detail.

\label{sect:emergece}
\begin{figure*}
\centering
\includegraphics[width=\textwidth]{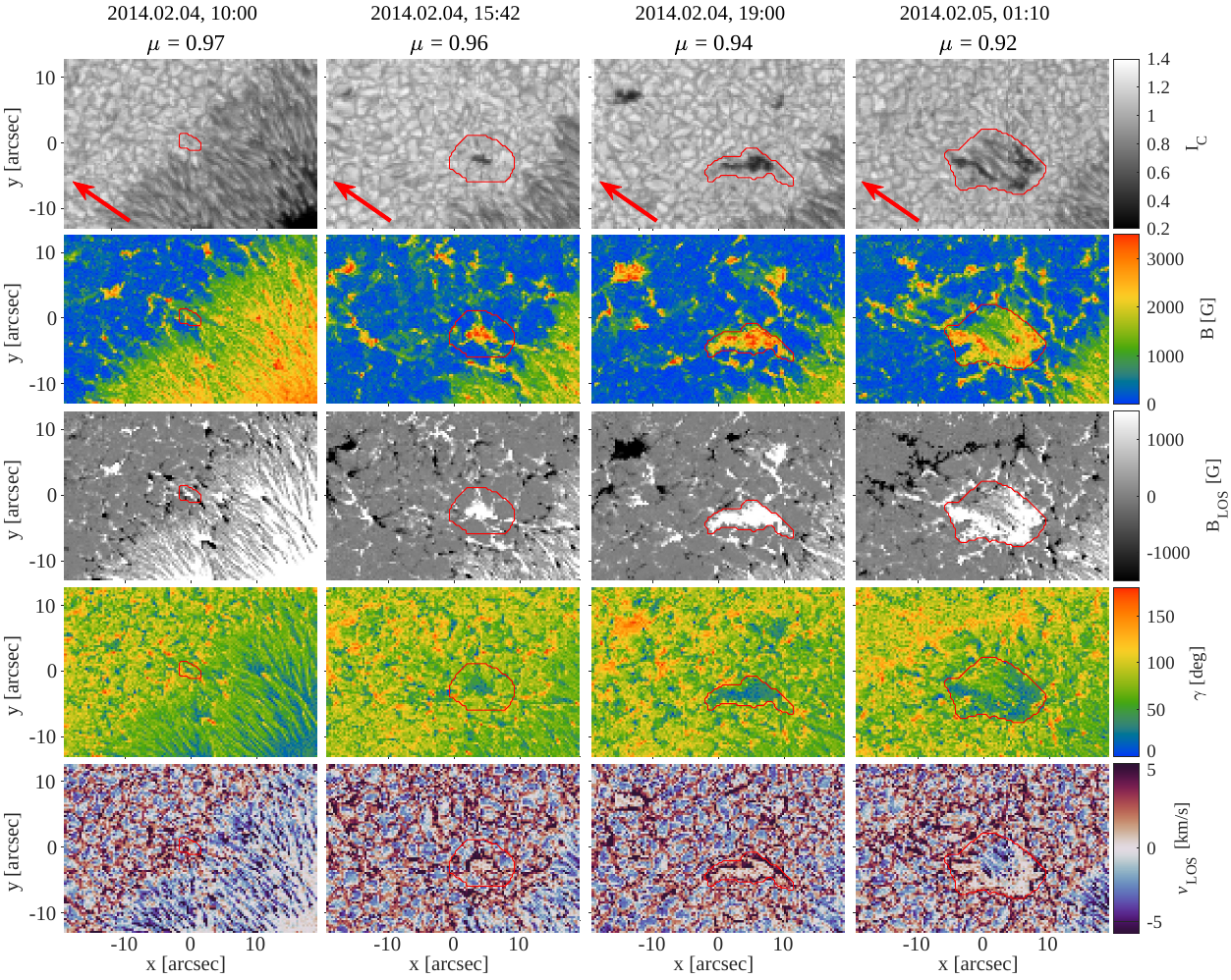}
\caption{Time-series showing the formation of an OPU in AR\,11967. The individual columns show the observables of the OPU at different times, from 10:00\,UT on 4 February 2014 to 1:10\,UT on 5 February 2014. {\it From top to bottom:} $I_{\rm c}$, $B$, $B_{\rm LOS}$, $\gamma$, and $v_{\rm LOS}$. All panels have a FOV of $38\farcs4 \times 25\farcs6$. The magnetic feature that evolves into an OPU (and the OPU itself) is indicated by the red contours. This OPU forms by emerging in a plage region. We show the magnetic field and the velocity at $\tau = 1$. The arrows in the top row point towards disk centre.}
\label{fig:OPU_emergence}
\end{figure*}

Figure~\ref{fig:OPU_separation} shows an example of an OPU that formed by separating from the penumbra of a sunspot (an OPU in AR\,11158 on 16 February 2011). Over the course of about $6.5$\,hr (from 1:00 UT to 7:30 UT), a patch of the penumbra of the sunspot detached from the spot, turning into an OPU. As can be seen in Fig.~\ref{fig:OPU_separation}, the filaments in the OPU did not exhibit any obvious changes while being detached from the spot, although the size of the detached part of the penumbra decreased with time. Also visible is that a small part of the outer edge of the umbra came along with the detaching penumbra and remained attached to the OPU as small pores. While detaching from the sunspot, the OPU carried a significant amount of magnetic flux away from the spot. During the OPU breakaway, the umbra stayed without a penumbra for 3\,hr, during which the umbra reduced its area. Subsequently, the penumbra regrew in the vacated region, covering the entire umbral region in a process that lasted another 3\,hr. At the time of the OPU breakaway event, the sunspot group was already decaying, suggesting that the OPU breakaway event might be related to the decay process of the group. 
A similar process was described by \citep{Jurcak2017A&A...Pore}, who studied the vertical magnetic field at the boundary of a pore. The authors proposed that the pore did not fulfil a threshold of $\sim$1.8\,kG \citep[][cf. \citet{2020A&A...639A.106L}]{2011A&A...531A.118J} and unusually, penumbrae eventually colonized the pore.

Note that not all OPUs detach accompanied by part of an umbra. Some OPUs separate only as penumbral filaments without umbrae associated with them.

\subsection{Emergence in an AR plage region} 

The second major mechanism for the formation of OPUs is the emergence of magnetic flux in a plage region. Such OPUs tend to form close to the PIL of the AR. Here we discuss this formation process using an OPU in AR\,11967 as an example. This OPU forms in a plage region close to a sunspot between 10:00\,UT and 22:00\,UT on 4~February 2014. The formation process of this OPU was captured by a series of Hinode/SOT-SP scans (we show the inverted maps in Fig.~\ref{fig:OPU_emergence}).  However, the spatial resolution of the HMI data is significantly lower than that of the Hinode/SOT-SP data. Hence, in the following, we will refer to the HMI observations to describe the time dependence of the large-scale formation process while adding information about the small spatial scales derived from the Hinode/SOT-SP data.

The formation process of the OPU starts with a small relatively vertical magnetic feature at the outer edge of the penumbra of the sunspot. It consists of mainly positive polarity field and is located close to the outer boundary of the sunspot's penumbra (see the left panel in Fig.~\ref{fig:OPU_emergence}). The temporal evolution of the region shows that the magnetic element detaches from the outer boundary of the sunspot around 10\,UT (see first column of Fig.~\ref{fig:OPU_emergence}) and then moves away from the penumbral region where it originated.

After separating from the sunspot, the magnetic flux increases in the magnetic element, leading to the formation of a small pore at around 13:18\,UT. In the subsequent hours, the size and the total magnetic flux of the pore increase. At 19:43\,UT and 20:16\,UT, two additional pores having positive polarity become visible in close vicinity of the existing pore. Opposite polarity patches are also seen in the HMI magnetograms. Still, they appear sparse and do not form a clear structure in the continuum images, at least at the HMI's spatial resolution. Over the course of the next two hours, the three pores merge. None of these pores shows any peculiarities. The merger is completed at around 22:00\,UT. While the pores were merging, the OPU started forming and reached its maximum area three hours later. The HMI magnetograms and continuun images reveal that the formation of the OPU is related to the emergence of fresh magnetic flux. Elongated convective cells, as well as the emergence and later expansion of the bipolar region, are typical signs of it. Then, the OPU slowly starts disappearing in a lapse of ten hours. The leftovers were again small pores that decayed in the moat of the main spot. 

The southern and western edges of the OPU consist of stronger, more vertical magnetic fields, resembling the ones of pores or spines. These regions do not exhibit significant line-of-sight flows. However, the outer edges of these pore-like structures and of the pore that precedes the OPU harbour strong downflows, reaching velocities of up to 13\,km/s in the bottom node. Strong downflows at the outer edges  occur for most pores \citep[e.g.,][]{1999ApJ...510..422K, Jafarzadeh2024A&A...PoreswavesPHI, Peng2024ApJ...StatisticsPores, Verma2024A&A...PoresGREGOR}. Within the OPUs and sunspot penumbrae, downflows are localized at the tails of individual OPU filaments. The tails also harbour magnetic fields of the opposite polarity to the head.

\begin{figure*}
\centering
\includegraphics[width=.85\textwidth]{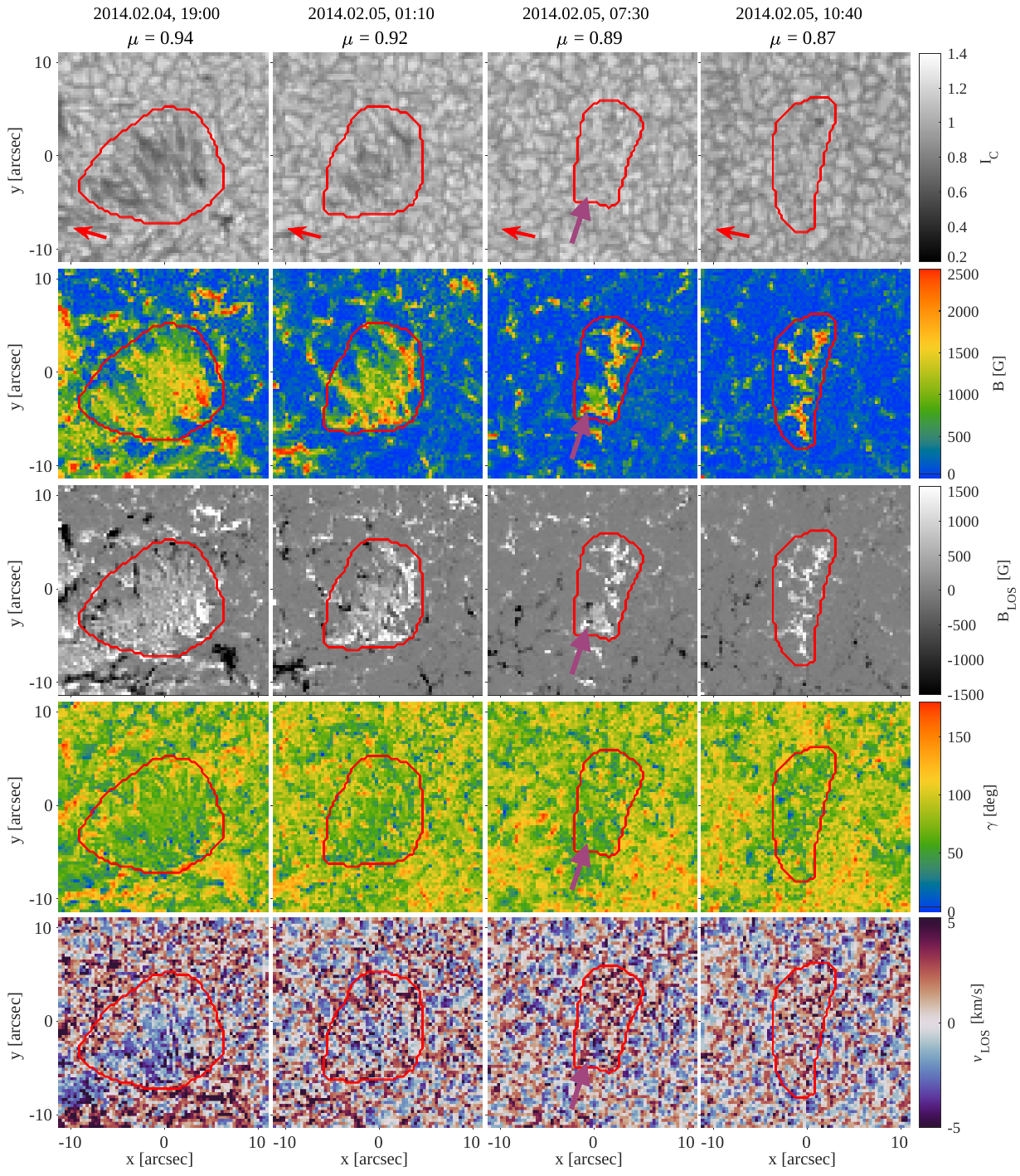}

\caption{Time-series showing the decay of an OPU in AR\,11967. The individual columns show the observables of the OPU at different times, from 19:00\,UT on 4 February 2014 to 10:40\,UT on 5 February 2014. {\it From top to bottom:} $I_{\rm c}$, $B$, $B_{\rm LOS}$, $\gamma$, and $v_{\rm LOS}$. All panels have a FOV of $22\farcs4 \times 22\farcs4$ and the red contours indicate the orphan penumbra and its remains. We show the magnetic field and the velocity at $\tau = 1$. The red arrows in the top row point towards disk centre. 
The purple arrows in the third column mark the leftover magnetic flux concentration.}
\label{fig:OPU_decay}
\end{figure*}

The moat flow is observed outside sunspots and parallel to the penumbral filaments in the locality \citep{Sheeley1972SoPh...moatflow, Harvey1973SoPh...MMF, VargasDominguez2007ApJL...moatflow, Strecker2018A&A...moatflow}. Although the OPU is located on the moat of the main spot, the HMI data also show the onset of a moat flow outside the OPU contemporaneous with the formation of filaments within the OPU. We could disentangle the moat flows of the spot and the OPU, as in the locality of OPU, the flow is parallel to the OPU filaments and not to the spot's filaments.

\section{Decay of orphan penumbrae}

\begin{figure*}
\centering
\includegraphics[width=.93\textwidth]{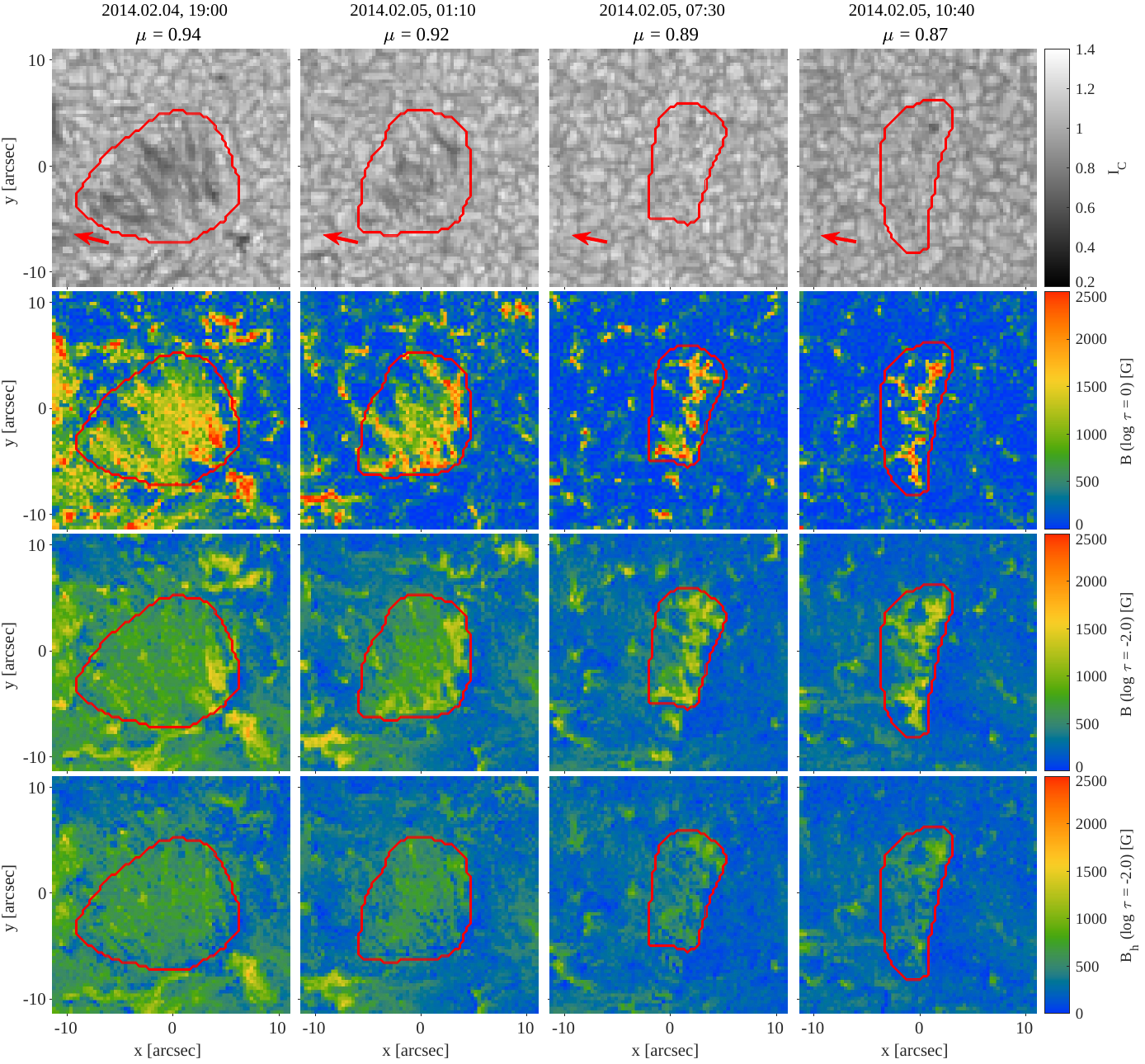}
\caption{Time-series of the magnetic field of the OPU in AR\,11967 (the one shown in Fig.~\ref{fig:OPU_decay}) during its decay process. The individual columns show the observables of the OPU at different times, from 19:00\,UT on 4 February 2014 to 10:40\,UT on 5 February 2014. {\it From top to bottom:} $I_{\rm c}$, $B$ at $\log\tau~=~0$, $B$ at $\log \tau = -2.0$, and strength of the horizontal magnetic field at $\log \tau = -2.0$. All panels have a FOV of $22\farcs4 \times 22\farcs4$ and the red contours indicate the orphan penumbra and its remains. The arrows in the top row point towards disk centre.}
\label{fig:OPU_decay_B}
\end{figure*}

The OPUs in the \modest{} catalogue have a typical lifetime of a few hours (up to 36\,h), after which they decay. Here, we discuss the decay process of OPUs using an OPU in AR\,11976 as an example (a different OPU to the one discussed in Sect.~\ref{sect:emergece}). The decay of this OPU is representative of the sample of the OPUs in the \modest{} catalogue. This OPU decayed over the course of several hours, from about 19:00\,UT on 4 February 2014 to about 6:00\,UT on 5 February 2014. During this time the OPU was observed multiple times by Hinode/SOT-SP. Figure~\ref{fig:OPU_decay} shows the time-evolution of various observables derived from the Hinode/SOT-SP scans at $\tau = 1$ and Fig.~\ref{fig:OPU_decay_B} shows the evolution of the magnetic field of the OPU at different optical depths. 

The OPU studied here does not exhibit any peculiarities before the decay process sets in (see the maps obtained at 19:00 on 4~February in Figs.~\ref{fig:OPU_decay} and \ref{fig:OPU_decay_B}). Over the next few hours, the OPU slowly decays. At 1:10\,UT on 5 February, the OPU has decreased in area (from $\sim$77\,Mm$^2$ at 19:00\,UT on 4~February to $\sim$57\,Mm$^2$). In addition, the OPU is not a homogeneous structure any more. Granulation has protruded into the OPU, in between individual filaments. This process also affects the magnetic field. At $\tau = 1$, the regions covered by granulation are almost field-free. However, there is still a strong horizontal field at $\log \tau = -2.0$ above the granulation and in the vicinity of the OPU. At lower levels in the atmosphere, this field is weaker, clearly indicating that it is a canopy field. 

\begin{figure*}
\centering
\includegraphics[width=\textwidth,trim={0 .65cm 0 .8cm},clip]{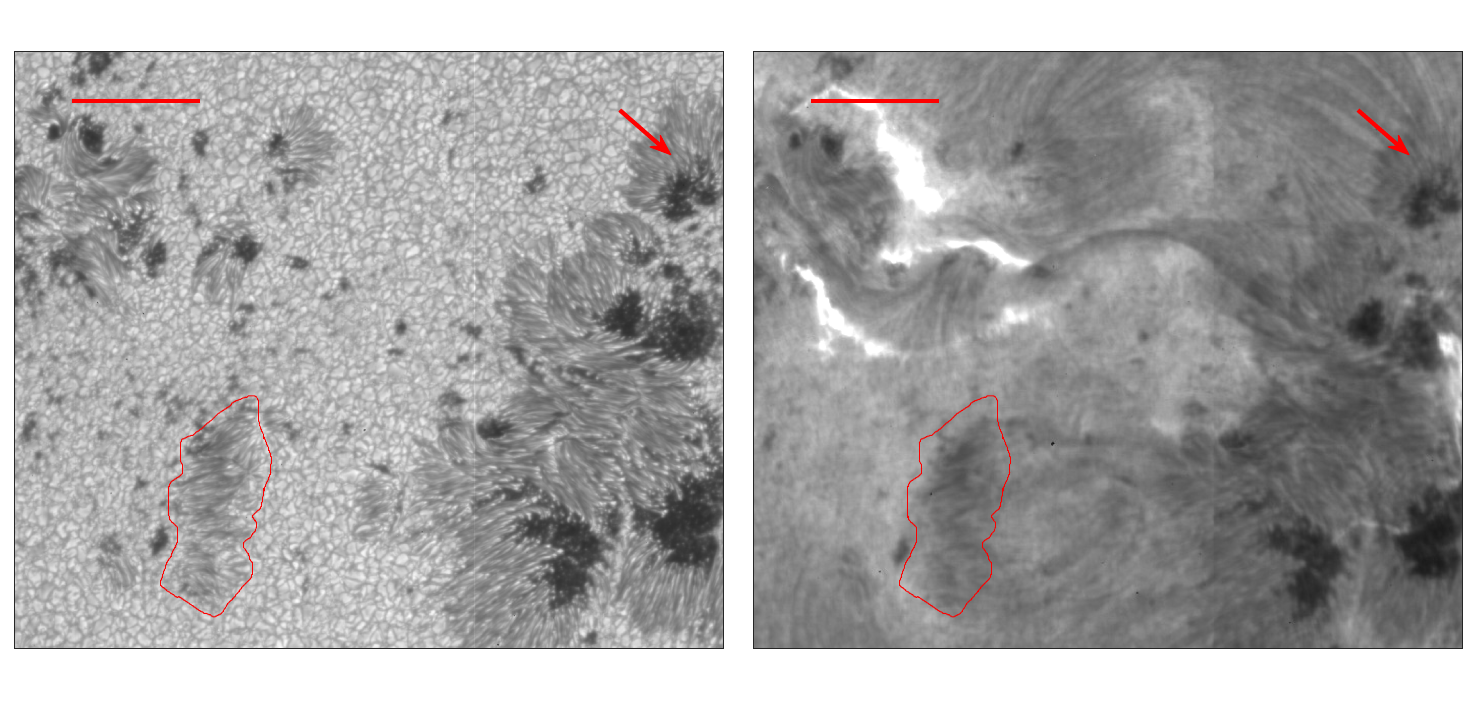}
\includegraphics[width=\textwidth,trim={0 .16cm 0 .1cm},clip]{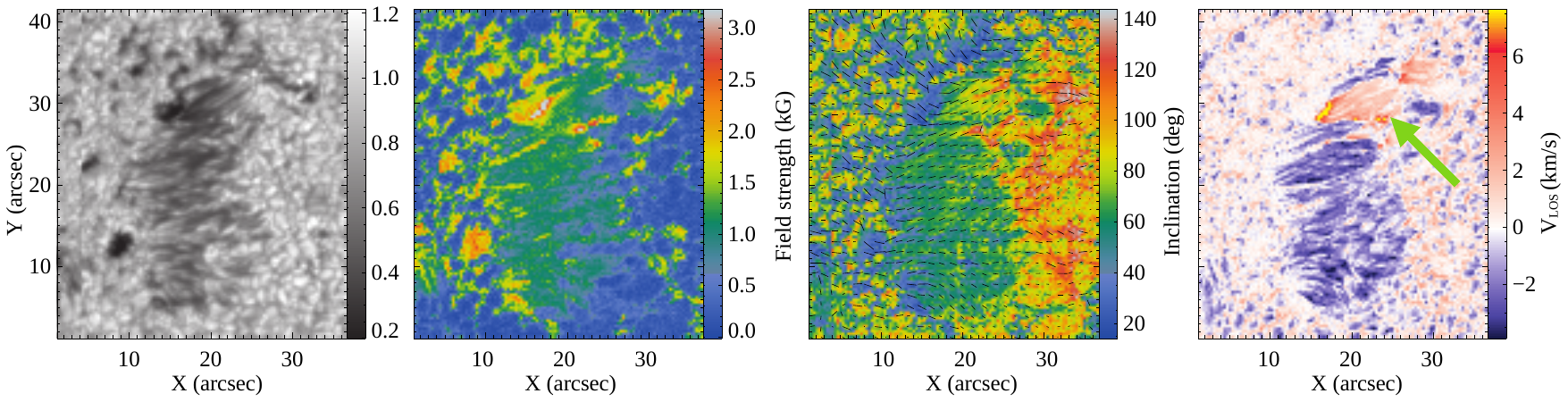}
\caption{\textit{Top row:} Maps of the continuum intensity (left) and of the H$\alpha$ line core intensity (right) of AR\,11339 on 6 November 2011 obtained by Hinode/SOT-FG. The red contour highlights the position of an OPU. The horizontal line indicates a length of $20\arcsec$ and the arrow points towards disk centre. The AR was located at a heliocentric angle $\mu = 0.89$. \textit{Botton row:} Panels displayed from left to right show the maps of the OPU of the continuum, magnetic field strength, inclination and azimuth (black lines), and LOS velocity.}
\label{fig:OPU_Halpha}
\end{figure*}

During the decay process, the OPU rotates in the counterclockwise direction and the magnetic field becomes elongated in the north-south direction. The decay of the OPU then starts at its northern edge. The decay process consists of individual penumbral OPU filaments disappearing and being replaced by granulation. The signature of the OPU filaments in the continuum intensity disappears at the same time. The magnetic field in this region gets more compact and gets concentrated at the western edge, where the heads of the OPU filaments used to be located. Simultaneously with the decay of the OPU filaments, the moat flow also ceases in these regions. This process stretches over the course of several hours, until the last remaining OPU filaments decay at around 6:00\,UT on 5 February.

The remains of the OPU consist of a region of enhanced magnetic field strength (see third column in Figs.~\ref{fig:OPU_decay} and~\ref{fig:OPU_decay_B}, at 7:30\,UT on 5~February). It resembles a typical magnetic flux concentration in plage areas (a filigree-like structure), with the magnetic field being concentrated in the downflowing intergranular lanes. There is still a canopy field extending to the east from the magnetic flux concentration. Most of the leftover magnetic flux concentration is covered by granulation (purple arrows in Fig.~\ref{fig:OPU_decay}). However, there is a region of enhanced magnetic field (up to $\sim$1000\,G) in the south of the feature, which has about the size of a granule and exhibits an upflow. It is surrounded by a ring of downflows in which the magnetic field has opposite polarity. Such a configuration of flows and magnetic field is untypical for plage regions but resembles sunspot penumbral filaments. Hence, this feature may be a remainder of the decayed OPU. Unfortunately, the spatial resolution of HMI is too low to resolve this feature and so, we cannot trace it back in time. At the time of the next Hinode/SOT-SP scan (at 10:40\,UT), this feature has disappeared and the magnetic flux concentration that remains from the OPU does not exhibit any differences from other magnetic elements in plage regions. The canopy field at $\log \tau = -2.0$ has decayed as well, although its remnants are is still visible (bottom right panel of Fig.~\ref{fig:OPU_decay_B}).

\section{Chromospheric counterpart of orphan penumbrae}

Several studies found a connection between OPUs and overlying magnetic fields in the lower chromosphere \citep[e.g.,][]{2012A&A...539A.131K,Kuckein2012A&A...FilamentOPUII,2013ApJ...769L..18L,2014ApJ...786L..22G,2014ApJ...787...57Z}. Measuring magnetic fields in the chromosphere is not straightforward. However, the magnetic field leads to the formation of structures in the chromosphere, such as fibrils, which can be observed in, e.g., H$\alpha$ images. The connection between such features and OPUs is not clear yet due to the small sample of OPUs that has been analysed so far. In this section, we will address this question by combining the observations of OPUs in the \modest{} catalogue with H$\alpha$ images provided by the Hinode/SOT-FG. 25 OPUs from the \modest{} catalogue were also observed in H$\alpha$. Unfortunately, Hinode/SOT-FG did not observe the OPUs in H$\alpha$ continuously, but only at selected times. The H$\alpha$ data does not cover the formation or decay process of any of the OPUs in our sample. Nonetheless, we can evaluate whether the presence of an OPU is related to the existence of structures such as filaments in the chromosphere.

Isolated OPUs are connected to fibrils (see Fig.~\ref{fig:OPU_Halpha} for an example). The footpoints of the fibrils are located within the OPU. They are aligned along the direction of the filaments within the OPU and they extend outwards from the OPU. They connect the OPU with patches of opposite polarity. This geometry indicates that the fibrils represent the continuation of the magnetic field of the OPU to higher atmospheric layers. If this is the case, the magnetic field in the chromosphere above the OPU is probably highly inclined. Hence, it could act as a barrier to the magnetic field in the photosphere, keeping it close to horizontal. However, it is also possible that the fibrils are merely a consequence of the existence of the OPU. Distinguishing between these two scenarios would require a time series of H$\alpha$ images capturing the formation of an OPU. 

The closest MODEST inversion of the OPU is shown in the bottom row of Fig.~\ref{fig:OPU_Halpha}. Both magnetic field information and LOS velocity are similar to those observed in typical OPUs, suggesting that this region is a typical OPU.  The magnetic field azimuth is shown with the black lines on the third panel of Fig.~\ref{fig:OPU_Halpha}. The azimuth along the OPU filaments aligns well with the H$\alpha$ broadband image. However, it is difficult to disentangle whether this is due to an alignment between the photospheric and chromospheric magnetic field of the OPU or photospheric contamination of the  H$\alpha$ broadband image. Distinguishing between these two scenarios would require co-temporal measurements of the photospheric and chromospheric magnetic fields.

Interestingly, most of the OPU shows blue-shifted Evershed-like flow, but on the northern part of the OPU, the LOS velocity along the OPU's filaments shows a patch with a reversed direction of the Evershed-like flow (green arrow). The flow direction at the patch is directed towards a dark region, resembling a micro-umbra. The magnetic field strength in the micro-umbra is about 2.5--3\,kG, which is typically found in pores and in not-too-dark umbrae. The continuum image shows no difference in the OPU filaments of the regions harbouring opposite flow directions, similar to the so-called counter Evershed flows \citep{2014A&A...564A..91J, 2014ApJ...787...57Z, Siu-Tapia2017A&A, Siu-Tapia2018ApJ, CastellanosDuran2021...rareCEFs, CastellanosDuran2023...ejectionCEFs}.

We note that fibrils are also present around sunspots (see Fig.~\ref{fig:OPU_Halpha}, top row). However, interpreting the Hinode/SOT-FG broadband H$\alpha$ line core images is challenging. The broadband mixes chromospheric and photospheric signals. The visibility of photospheric features, such as pores, umbrae, and the darkened penumbra, suggests a significant photospheric contribution. While fibrils are apparent, their classification as chromospheric or photospheric structures remains ambiguous, particularly within the penumbra and OPUs. These features may represent photospheric filaments observed through the H$\alpha$ filter. However, the curved dark feature ending in one of the flare ribbons in Fig.~\ref{fig:OPU_Halpha} is an actual filament, i.e. projection onto the disc of a prominence. The narrow filaments ending in the OPU look to be low-lying chromospheric loops. The filament touches the flare ribbon and lies on top of a polarity inversion line. Also, the OPU has the opposite polarity to the sunspot, and fibrils start in the spot and end in the OPU.

Some previous authors attributed the formation of at least some OPUs to low-lying fibrils. We do not see any OPU in the \modest{} sample, which is related to fibrils, apart from the OPU in AR\,10953 \citep[which was studied by][]{2016A&A...589A..31B}. Hence, fibrils do not seem to appear to play a major role in the origin of OPUs.

\section{Discussion and conclusions}
Orphan penumbrae are remarkably similar, irrespective of their differences in size and formation process. The filaments of OPUs are also remarkably similar to penumbral filaments. However, it seems counterintuitive since OPUs display a broad range of shapes and partly different configurations of the magnetic field. Most OPUs are dominated by one polarity of the magnetic field. In some cases, however, there is a strong influence of the opposite polarity, with the OPU connecting the two magnetic polarities. Despite these differences in the structure of OPUs, the OPU filaments are remarkably similar between different OPUs, resembling in important aspects the ones in sunspots. The scatter plots in Fig.~\ref{fig:OPU_example_scatter} suggest that only the brightness of the OPU filaments varies between different OPUs. This further supports the results of \citet{2013A&A...557A..25T} and \citet{2021A&A...655A..61L} about the uniformity of the properties of penumbral filaments. The uniformity of penumbral filaments is remarkable given the broad range of environments they are embedded in, now including OPUs. Looking beyond individual filaments, there are  differences between OPUs and sunspot penumbrae. In sunspots, the penumbral filaments are interlaced with spines, in which the magnetic field is stronger and more vertical. Orphan penumbrae, however, do not exhibit spines \citep[cf.][]{2014A&A...564A..91J}. Given that spines are like an extension of the umbra into the penumbra, we expect that the absence of spines in OPUs is simply a result of the absence of a bordering umbra, and the subsequent difference in the magnetic field structure in sub-surface layers. Likewise, the outer penumbrae of spots harbours few or no spines, suggesting the sunspot penumbra's global structure also plays a role, which is not the case for OPUs.  

We identified two major mechanisms for the formation of OPUs: the separation from the host sunspot's penumbra and the emergence of new magnetic flux. Orphan penumbrae that form when a part of a sunspot's penumbra gets detached tend to stay in close proximity to the spot. Hence, such OPUs are still covered by the canopy field of the sunspot. The canopy probably keeps the magnetic field from rising to higher atmospheric layers, causing it to be strongly inclined. Orphan penumbrae that form due to the emergence of new flux are typically located in the moat of spots and close to the PIL of the ARs. Hence, the magnetic field has a flat, very low-lying shape of $\Omega$ loops, connecting the OPUs with magnetic elements of opposite polarity nearby, in agreement with \citet{2014A&A...564A..91J}.  A somewhat different geometry of the magnetic field is suggested by H$\alpha$ images, which show fibrils fanning out from the OPU across the PIL, as this suggests that the field has a single polarity and is not connecting opposite magnetic polarities at the two ends of an OPU. Taken together, the two observations suggest that, like in penumbrae, a part of the field continues in the form of a magnetic canopy beyond the OPU \citep{Solanki1994A&A...Infrared-lines..Evershed}, while another part of the field returns below the solar surface at the outer edge of the OPU \citep[e.g.,][]{WestendorpPlaza1997Natur...Evershed}.

The LOS velocity map of the OPU part of AR 11339 shows a region with an opposite direction of the Eveshed-like flow along the OPU compared to other parts. These counter Evershed-like flows in OPU are similar to those found in OPUs belonging to AR 10960  investigated by \citet{2014A&A...564A..91J}, AR 11089 \citep{2014ApJ...787...57Z}, and in AR 12674  \citep[][see Fig.~1.3]{CastellanosDuran2022...Phd}. A recent statistical study found that regular penumbral filaments associated with an umbra harbour normal Evershed flow for 94\% of the time they are observed, and the rest of the time, penumbral filaments harbour counter Evershed flows \citep{CastellanosDuran2021...rareCEFs}. 

It is now widely agreed that normal Evershed flow in sunspot penumbrae is the result from overturning magnetoconvection due to the overlaying magnetic field \citep{2008ApJ...677L.149S, 2011ApJ...729....5R, 2013A&A...557A..25T}. However, the mechanism driving counter-Evershed flows is still under debate \citep[e.g.,][]{Siu-Tapia2018ApJ, 2024A&A...686A.112G}. As OPUs are not associated with umbrae, the spot's magnetic field imposes no direction of preference, therefore, further analyses of the seemingly routine appearance of counter Evershed-like flows in OPU could help improve our understanding of the physical conditions that drive the flows in OPU as well as the counter Evershed flows along regular penumbral filaments. 

The Hinode observations of the decay of the OPU in AR\,11967 show that the canopy field in the upper photosphere is still present after the OPU has decayed (see Fig.~\ref{fig:OPU_decay_B}). Such behaviour was also reported by \citet{2016A&A...589A..31B}. This suggests that the decay starts in deeper layers, probably driven by the turbulent convection at and below the solar surface. With time the field in the middle of the OPU rises. Once it is mainly above the solar surface, the OPU ceases to exist as a single photometric structure visible in continuum intensity. Nonetheless, magnetic field at its two ends still endures for some more time, connected by a magnetic canopy. 

While the results discussed above indicate a connection between the formation of OPUs and a nearly horizontal magnetic field in the lower chromosphere and upper photosphere, there are still uncertainties on how exactly this process works. A clearer picture of the formation of OPUs requires a simultaneous time series of observations of an OPU in both the photosphere and in the chromosphere with high spatial and temporal resolution. Such observations were possible with  the Sunrise~III mission \citep{Korpi-Lagg2025...SunriseIII}. Also, the novel MiHI instrument \citep{vanNoort2022A&A...MihiInstrument}, which provides integral field spectropolarimetry observations, observed an OPU. The analysis of Sunrise~III and MiHI observations could provide unique simultaneous information from the temporal, spatial, and spectral dimensions of OPUs.

\begin{acknowledgements}
We thank the referee, Dr.~Jan~Jur{\v{c}}{\'a}k, for insightful comments that improved the quality of the manuscript. This project has received funding from the European Research Council (ERC) under the European Union’s Horizon 2020 research and innovation programme (grant agreement No. 101097844 — project WINSUN). Hinode is a Japanese mission developed and launched by ISAS/JAXA, collaborating with NAOJ as a domestic partner, and NASA and STFC (UK) as international partners. Scientific operation of the Hinode mission is conducted by the Hinode science team organized at ISAS/JAXA. This team mainly consists of scientists from institutes in partner countries. Support for the post-launch operation is provided by JAXA and NAOJ (Japan), STFC (U.K.), NASA, ESA, and NSC (Norway). The data were processed at the German Data Center for SDO (GDC-SDO), funded by the German Aerospace Center (DLR). The HMI data used are courtesy of NASA/SDO and the HMI science team.
\end{acknowledgements}

\bibliographystyle{aa} 
\bibliography{literature}

\end{document}